\documentclass[preprint,showpacs,preprintnumbers,amsmath,amssymb,nofootinbib]{revtex4}
\usepackage{booktabs}
\usepackage{mathrsfs}
\usepackage{epsfig}
\usepackage{graphicx}
\usepackage{dcolumn}
\usepackage{bm}
\usepackage{amsmath}
\usepackage{slashed}
\usepackage{multirow}

\let\jnfont=\rm
\def\NPB#1,{{\jnfont Nucl.\ Phys.\ B }{\bf #1},}
\def\PLB#1,{{\jnfont Phys.\ Lett.\ B }{\bf #1},}
\def\EPJC#1,{{\jnfont Eur.\ Phys.\ Jour.\ C }{\bf #1},}
\def\PRD#1,{{\jnfont Phys.\ Rev.\ D }{\bf #1},}
\def\PRL#1,{{\jnfont Phys.\ Rev.\ Lett.\ }{\bf #1},}
\def\MPLA#1,{{\jnfont Mod.\ Phys.\ Lett.\ A }{\bf #1},}
\def\JPG#1,{{\jnfont J.\ Phys.\ G}{\bf #1},}
\def\CTP#1,{{\jnfont Commun.\ Theor.\ Phys.\ }{\bf #1},}
\def\ZPC#1,{{\jnfont Z.\ Phys.\ C }{\bf #1},}
\def\JHEP#1,{{\jnfont JHEP \ }{\bf #1},}
\def\Rv{\not{\hbox{\kern-1pt $R$}}}
\def\p{\not{\hbox{\kern-3pt $p$}}}
\def\GeV{\text{ GeV}}
\def\MeV{\text{ MeV}}

\newcommand{\bea}{\begin{eqnarray}}
\newcommand{\eea}{\end{eqnarray}}

\newcommand{\bcen}{\begin{center}}
\newcommand{\ecen}{\end{center}}

\newcommand{\beq}{\begin{eqnarray}}
\newcommand{\eeq}{\end{eqnarray}}

\def\t1{\tilde{t_1}}

\begin{document}

\title{An indirect probe of higgsino world at the CEPC}
\author{ Ning Liu$^1$}
\author{ Lei Wu$^2$}
\affiliation{
$^1$ Institution of Theoretical Physics, Henan Normal University, Xinxiang 453007, China\\
$^2$ Department of Physics and Institute of Theoretical Physics, Nanjing Normal University, Nanjing, Jiangsu 210023, China
}%


\begin{abstract}
The higgsino world is one of the popular natural SUSY scenarios, in which $|\mu| \ll M_{gauginos} \ll M_{scalars}$. As such, searching for light degenerate higgsinos becomes an essential task of testing naturalness in supersymmetry (SUSY). In this work, we study the indirect effects of light higgsinos in the process $e^+e^- \to W^+W^-$ at future Higgs factories, such as Circular Electron Positron Collider (CEPC) in China. we find that the higgsino mass parameter $\mu \lesssim 210$ GeV favored by naturalness can be covered if the accuracy of the measurement of $W^+W^-$ production can reach 0.1\% at 240 GeV CEPC.
\end{abstract}

\maketitle

\section{INTRODUCTION}
In past decades, the naturalness (or hierarchy) problem of the Higgs mass has been the guiding principle to construct theories beyond the Standard Model (SM). Among various extensions, low-energy supersymmetry (SUSY) provides a framework for a light Higgs boson without invoking unnatural fine-tuning of theory parameters. Moreover, the criterion of naturalness has the strong implication that new dynamics should occur at a scale
around the TeV and be accessible to the current running LHC.

In the MSSM, the starting point of discussions of the naturalness is from the minimization of the tree-level Higgs potential, which leads to the following relation:
\begin{eqnarray}
\frac{M^2_{Z}}{2}&=&\frac{(m^2_{H_d}+\Sigma_{d})-(m^2_{H_u}+
\Sigma_{u})\tan^{2}\beta}{\tan^{2}\beta-1}-\mu^{2},
\label{minimization}
\end{eqnarray}
where $m^2_{H_{u,d}}$ denote the soft SUSY breaking masses of the Higgs fields at the weak scale, respectively. $\tan\beta = v_u/v_d$ and $\mu$ is the higgsino mass parameter. $\Sigma_{u,d}$ arise from the radiative corrections to the tree level Higgs potential  \cite{mz}. In order to get correct value of $M_Z$ without highly tuning the theory parameters, each term on the right-hand side of Eq.(\ref{minimization}) should be comparable in magnitude \cite{bg,baer-ew-0}. This indicates four light higgsinos in the spectrum. Additionally, the stop and gluino can contribute to $\Sigma_u$ through one-loop and two-loop radiative corrections and are expected to be less than about 600 GeV and 2 TeV, respectively, if the high scale $\Lambda=20$ TeV is assumed. Such a spectrum is typically predicted by the conventional natural SUSY \cite{nsusy-1,nsusy-2,nsusy-3} and has been widely studied \cite{stop-1,stop-2,stop-3,stop-6,stop-7,stop-8,stop-9,stop-10,stop-11,stop-13,stop-14,stop-15,stop-16,stop-18,stop-20,stop-22,stop-23,stop-25,stop-26}. However, the discovery of a 125 GeV SM-like Higgs boson \cite{higgs-atlas,higgs-cms} and the recent null results of the LHC Run-2 searches for sparticles have excluded the stop and gluino masses up to $\sim$ 1 TeV \cite{run2-stop} and 1.8 TeV \cite{run2-gluino}, respectively. Therefore, it is imperative to explore other possible scenarios for which the theory maintains naturalness.

The importance of $\mu$ parameter itself as a measure of fine-tuning was first emphasized by \cite{nath}, and then motivates the hyperbolic branch/ focus point (HB/FP) region of minimal supergravity (mSUGRA or CMSSM) \cite{nath,feng}, and recent radiative natural SUSY (RNS) \cite{baer-ew-0}, in which heavy scalars with low $\mu$ value and low fine-tuning are realized. In this paper, we will focus on such a kind of scenario in MSSM with large, multi-TeV scalar and gaugino masses, but with low, sub-TeV superpotential $\mu$ term,
\begin{eqnarray}
|\mu| \ll M_{gauginos} \ll M_{scalars},
\end{eqnarray}
which was named ``higgsino world'' by Kane \cite{kane} and can be easily realized in models with non-universal GUT scale Higgs masses (NUHM), such as the RNS. In the limit of $|\mu| \ll M_{gauginos}$, the two lightest neutralinos ($\tilde{\chi}^0_{1,2}$) and lighter chargino ($\tilde{\chi}^\pm_1$) are higgsino-like and are nearly mass degenerate, as shown below,
\begin{eqnarray}
 m_{\tilde{\chi}^\pm_1} - m_{\tilde{\chi}^0_1} &=&
  \frac{M_W^2}{2 M_2} \left( 1 - \sin 2\beta - \frac{2 \mu}{M_2} \right) + \frac{M_W^2}{2 M_1} \tan^2\theta_W (1 + \sin 2\beta) , \nonumber \\
m_{\tilde{\chi}^0_2} - m_{\tilde{\chi}^0_1} &=&
  \frac{M_W^2}{2 M_2} \left( 1 - \sin 2\beta + \frac{2 \mu}{M_2} \right) + \frac{M_W^2}{2 M_1} \tan^2\theta_W (1 - \sin 2\beta) \; .
\label{mass-difference}
\end{eqnarray}
Thus, the usual dilepton and trilepton signatures from electroweakino pair production are difficult to observe as the very soft $p_T$ spectrum of the isolated leptons in the final states at the LHC. One possible way is to use a hard ISR jet to trigger the events of electroweakino pair, however, this is challenging at high luminosity LHC (HL-LHC) due to the large systematical uncertainty \cite{higgsino-1,higgsino-2,higgsino-3,higgsino-4,higgsino-5}. On the other hand, the indirect searches for light higgsinos via quantum effects may paly an important and complementary role since their sensitivity depend on the mass splitting between higgisnos weakly.

The process $e^+e^- \to W^+W^-$ is one of the key processes at LEP2, and has been precisely calculated in the SM \cite{ww-sm-1,ww-sm-2,ww-sm-3,ww-sm-4,ww-sm-5,ww-sm-6}. Previous studies of the process $e^+e^- \to W^+W^-$ in supersymmetric theories include the complete one-loop corrections in spontaneously broken supersymmetry \cite{ww-mssm-1}, in the MSSM \cite{ww-mssm-2,ww-mssm-3,ww-mssm-4,ww-mssm-5}. The important contributions come from the sfermions, however, the electroweakinos can have  sizable contributions \cite{ww-mssm-4}, in particular considering the current strong LHC limits on sfermions. In this work, we investigate the indirect effects of higgsino world in the process $e^+e^- \to W^+W^-$ at 240 GeV CEPC, which will deliver 5 ab$^{-1}$ of integrated luminosity during 10 years of operation and can probe various new physics models with an unprecedented precision \cite{cepc-1,cepc-2,cepc-3,cepc-4,cepc-5,cepc-6,cepc-7,cepc-8,cepc-9,cepc-10,cepc-11,cepc-12,cepc-13,cepc-14,cepc-15}. At CEPC, about ${\cal O}(10^7)$ events of $W$ pair can be produced in a clean environment, which allows for the measurement of the cross section of $e^+e^- \to W^+W^-$ with ${\cal O}(0.1\%)$ precision. Therefore, the process $e^+e^- \to W^+W^-$ may be served as a probe of higgsino world in the MSSM. The structure of this paper is organized as follows. In Section \ref{section2}, we will give a description of the calculation. In Section \ref{section3}, I will scan the parameter space of MSSM and present the numerical results. Finally, we draw our conclusions in Section \ref{section4}.

\section{Calculations}\label{section2}
In the SM and MSSM, tree-level diagrams of the process $e^+e^- \to W^+ W^-$ are the same, which involve $\gamma$ and $Z$ exchange in the $s$- channel and neutrino exchange in the $t$-channel. The Higgs-exchange diagrams can safely be neglected because of the suppression of a factor $m_e/m_W$. For the designed energy of the CEPC, the tree-level cross section of $W$ pair production is dominated by the neutrino mediated transverse $W$ bosons. Since the sfermions and non-SM Higgs bosons are decoupled in higgsino world, the main MSSM corrections come from the $s$-channel through higgsino loops.

The higgsinos including two light neutralinos $\tilde{\chi}^0_{1,2}$ and two light chargino $\tilde{\chi}^\pm_{1}$ can contribute to the process $e^+e^- \to W^+ W^-$ at one-loop level. The couplings of the neutralinos and charginos to $W$ and $Z$ bosons take the form $ig\gamma^\mu[G_L P_L+G_RP_R]$, where $P_L=(1-\gamma_5)/2$ and $P_R=(1+\gamma_5)/2$. Without CP violation, $G_L$ and $G_R$ are:
\begin{eqnarray}
W^+ \tilde{\chi}^+_i \tilde{\chi}^0_j:&& \quad
G_L=-{1\over \sqrt 2}V_{i2}N_{j4}+V_{i1}N_{j2}\,,\quad
G_R=+{1\over \sqrt 2}U_{i2}N_{j3}+U_{i1}N_{j2}\,,\\
Z \tilde{\chi}^+_i \tilde{\chi}^-_j:&&\quad
G_L=V_{i1}V_{j1}+{1\over 2} V_{i2}V_{j2}\,,\quad
G_R=U_{i1}U_{j1}+{1\over 2} U_{i2}U_{j2}\,,\\\
Z \tilde{\chi}^0_i \tilde{\chi}^0_j:&&\quad
G_R=-G_L={1\over 2}\left(N_{i3}N_{j3}-N_{i4}N_{j4}\right)\,,
\end{eqnarray}
where the $N$ matrix diagonalizes the neutralino mass matrix (with $1,2,3,4$ referring to the $ B, W^0, H_1^0, H_2^0$ basis) and the $V$ and $U$ matrices diagonalize the chargino mass matrix (with $1,2$ referring to the $ W^\pm, H^{\pm}$ basis). It can be seen that when $|\mu|\ll M_{1,2}$, one has: $V_{11},U_{11}\sim 0$, $V_{12},U_{12}\sim {\rm
sign}(\mu),1$; $N_{11},N_{12},N_{21},N_{22}\sim 0$, $N_{13}=N_{14}=N_{23}=-N_{24}=1/\sqrt 2$. In this case, the $Z,\gamma \tilde{\chi}^+_1\tilde{\chi}^-_1$, $Z \tilde{\chi}^0_1 \tilde{\chi}^0_2$, $W^\pm \tilde{\chi}^+_1 \tilde{\chi}^-_1$, and $W^\pm \tilde{\chi}^+_1 \tilde{\chi}^0_2$ rates will be large and $Z \tilde{\chi}^0_1 \tilde{\chi}^0_1$, $Z\tilde{\chi}^0_2 \tilde{\chi}^0_2$ are suppressed.

We calculate the ${\cal O}(\alpha)$ radiative corrections to the process $e^+e^- \to W^+ W^-$ in 't~Hooft--Feynman gauge. We use the dimensional regularization to isolate the ultraviolet divergences (UV) in the one-loop amplitudes and remove the UV singularities by using the on-mass-shell renormalization scheme \cite{on-shell}. Since there are no sparticles contributing to the process $e^+e^- \to W^+ W^-$ at tree level, the counter-terms in MSSM are the same as those in the SM.

Besides UV divergences, the infrared (IR) divergences can appear in the virtual correction because of the exchange of virtual photon in the loops. These IR divergences can be canceled by including the real photon bremsstrahlung corrections. We regularize these IR divergences by an infinitesimal photon mass $\lambda$. We denote the momentums of initial and final states for the real photon emission process as follows:
\begin{equation}
e^+(p_1)+e^-(p_2) \to W^+(k_1)+W^-(k_2)+\gamma(k) \, .
\end{equation}
In the soft photon approximation \cite{soft-photon}, the cross-section for real photon emission, $ \Delta\sigma_{soft}$, is given by:
\begin{eqnarray}
\label{s} {d} \Delta\sigma_{soft} = d \sigma_{{0}}
\frac{\alpha}{2 \pi^2} \int_{E_{\gamma} \leq \Delta E_{\gamma}}
\frac{d^3 \vec{k}}{2 E_{\gamma}} \left( \frac{k_1}{k_1 \cdot k} -
\frac{k_2}{k_2 \cdot k} \right)^2,
\end{eqnarray}
where $E_{\gamma} = \sqrt{|\vec{k}|^2+\lambda^2}$ and $\Delta E_\gamma$ is the energy
cutoff of the soft photon and assumed to be $\Delta E_\gamma=0.05\sqrt{s}$. The hard photon ($E_\gamma \geq \Delta E_\gamma$) corrections can be directly evaluated by the numerical Monte Carlo method \cite{vegas}. We numerically checked the cancellation of the IR divergences by varying the photon-mass $\lambda$ from 1 to $10^{10}$ GeV and found our total cross section is almost independent of $\lambda$.

\section{Numerical Results and Discussions}\label{section3}
We generate the Feynman diagrams with \textsf{FeynArts} \cite{feynarts}. The resulting amplitudes are algebraically simplified by \textsf{FormCalc} \cite{formcalc} and then converted to a Fortran code. The \textsf{LoopTools} package \cite{looptools} was used to evaluate the one-loop scalar and tensor integrals numerically. We take the input parameters of the SM as
\begin{equation}
\begin{aligned}
\alpha(m_Z)^{-1} &= 127.918,\quad &
M_Z &= 91.1867\GeV,\quad &
M_W &= 80.385\GeV, \\[.5ex]
m_e &= 0.51099907\MeV,\quad &
m_u &= 53.8\MeV,\quad &
m_d &= 53.8\MeV,\quad \\
m_\mu &= 105.658389\MeV, &
m_c &= 1.50\GeV, &
m_s &= 150\MeV, \\
m_\tau &= 1777\MeV, &
m_t &= 173.07\GeV, &
m_b &= 4.7\GeV.
\end{aligned}
\end{equation}

In the higgsino world, only $\mu$ parameter is light and other SUSY parameters are decoupled by taking heavy masses. To satisfy the 125 GeV Higgs mass within a 3 GeV uncertainty, we set the stop soft masses at 10 TeV and vary the stop trilinear parameter in the range $|A_{t}|<2$ TeV and $10<\tan\beta<50$. The slepton soft masses and the first two generation squark soft masses are assumed as 10 TeV. Inspired by the grand unification relation, we assume $M_1:M_2 = 1:2$ and take $M_1=1$ TeV. Given the LEP bounds, we vary the parameter $\mu$ in the range $100~{\rm GeV} <|\mu|< 300~{\rm GeV}$. To show the contributions of the higgsinos to the process $e^+e^- \to W^+ W^-$, we define the relative corrections $\delta(e^+e^- \to W^+ W^-)$ as,
\begin{eqnarray}
\delta(e^+e^- \to W^+ W^-)=\frac{\sigma^{MSSM}_{\rm one-loop}-\sigma^{SM}_{\rm one-loop}}{\sigma^{SM}_{\rm one-loop}}
\end{eqnarray}
where $\sigma^{MSSM,SM}_{\rm one-loop}$ denotes the one-loop corrected total cross section of $e^+e^- \to W^+ W^-$ in the MSSM and SM, respectively. Since the cross section of $e^+e^- \to W^+ W^-$ in the MSSM is not sensitive to other SUSY parameters, such as $\tan\beta$, we will show the dependence of the relative correction $\delta(e^+e^- \to W^+ W^-)$ on the higgsino mass parameter $\mu$ in the following. On the other hand, it is worth mentioning that the MSSM corrections to the cross section of $ e^+e^- \to W^+ W^-$ can change sizably when the sfermion masses become light. The contributions from non-SM Higgs bosons are much smaller than those of higgsinos and sfermions.

\begin{figure}[ht]
\centering
\includegraphics[width=5in,height=2.8in]{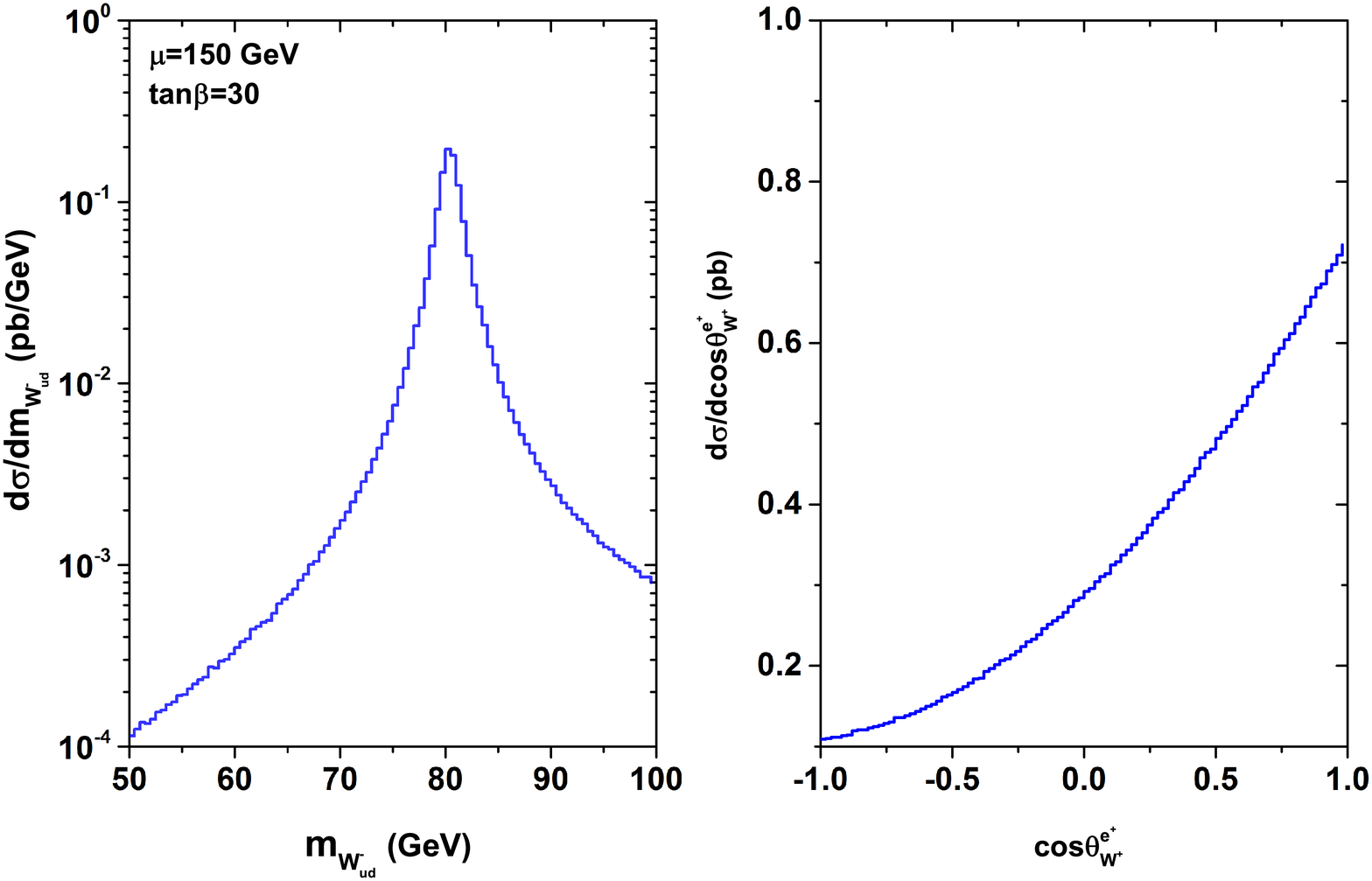}
\vspace{-0.5cm}
\caption{Parton-level kinematic distributions of the process $e^+e^- \to W^+( \to e^+\nu_e) W^-(\to \bar{u}d)$ at the CEPC. $m_{\bar{u}d}$ is the invariant mass of $\bar{u}$ and $d$ quarks from $W^-$ decay (left panel) and $\cos\theta^{e^+}_{W^+}$ is the cosine of the $e^+$ decay angle with respect to the $W^+$ direction in the CM system (right panel).}
\label{distributions}
\end{figure}
In Fig.~\ref{distributions}, we present the parton-level kinematic distributions of the process $e^+e^- \to W^+( \to e^+\nu_e) W^-(\to \bar{u}d)$ for $\mu=150$ GeV and $\tan\beta=30$ at the CEPC. From the Fig.~\ref{distributions}, we can see that the distribution of the invariant mass of $\bar{u}$ and $d$ quarks from $W^-$ decay shows a peak around the $m_{W}$. The anti-lepton $e^+$ from $W^+$ decay tends to fly along the $W^+$ boson direction in the CM system. We also find that these distributions are similar to the SM predictions. So, in order to probe the light higgsinos through process $ e^+e^- \to W^+W^-$, the SM background normalization and shapes should be known very well and the interplay with the best theoretical predictions (via MC) and data are necessary. At the CEPC, $W$-pair events are categorized and selected according to the decay products of $W$ bosons. For fully leptonic channel, both $W$¡¯s decay into a lepton-neutrino pair. This decay generates a low particle multiplicity and a large missing transverse momentum. Besides the two selected primary leptons will be energetic and have a large acoplanarity in the plane transverse to the beam line. Dominant background come from $e^+e^- \to \tau^+\tau^-$ processes. For the semi-leptonic channel, the primary lepton is energetic and isolated from the two jets of the hadronic system. The primary neutrino will produce a large missing momentum vector, also isolated in space. Main backgrounds originate from $Q\bar{Q}$ events with energetic leptons from heavy quark decays, and other four fermion processes, such as $e^+e^- \to ZZ$. For fully hadronic channel, four jet events without missing energy are characterized by a very large particle multiplicity with a spherical momentum distribution. The large background is from non-radiative $q\bar{q}$ events with hard QCD gluon emissions. Several observables are designed to efficiently suppress those backgrounds and the detectors are expected to be optimized, which may help to achieve the measurements of $W$ pair production with high precision at the CEPC~\cite{cepc-0,cepc-ew}.

\begin{figure}[ht]
\centering
\includegraphics[width=4in,height=2.8in]{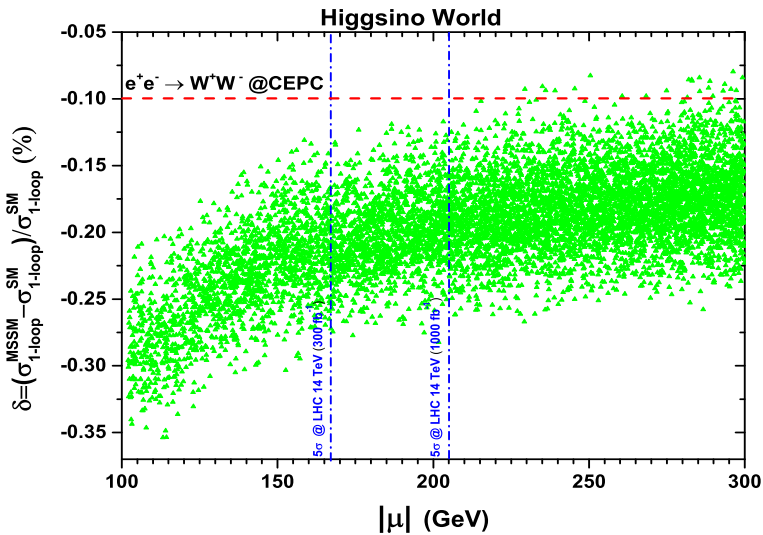}
\vspace{-0.5cm}
\caption{Scatter plot of relative one-loop corrections to the process $e^+e^- \to W^+ W^-$ in the higgsino world of MSSM at the CEPC. The blue vertical dash-dot lines are the expected $5\sigma$ sensitivity of LHC search for higgsino world via monojet plus soft dileptons plus $\slashed E_T$ \cite{higgsino-3}.}
\label{ratio}
\end{figure}
In Fig.~\ref{ratio}, we show the relative one-loop corrections to the process $e^+e^- \to W^+ W^-$ in the higgsino world of MSSM at the CEPC. The effect of currently measured $W$ boson mass on the cross section of $e^+e^- \to W^+W^-$ is negligible small, which is given in Tab.~\ref{mw}. We can see that  The measurement precision of $W$ mass is expected to be 3 MeV at CEPC, and the beam energy uncertainty can reach up to 10 ppm, $\sim 1$ MeV. The effects of radiative corrections and detector simulation are also well controlled \cite{cepc-0,cepc-ew}. In Ref.~\cite{cepc-12}, the authors estimate the main uncertainty in $W^+W^-$ measurement and found that the statistical errors are expected to dominate the $W^+W^-$ measurements at CEPC. The detailed analysis of the systematic errors needs a more realistic detector performances of CEPC, which is still not available. From Fig.~\ref{ratio}, we can see that the relative correction $\delta(e^+e^- \to W^+W^-)$ decrease with the increase of higgsino mass. The maximal value of $\delta(e^+e^- \to W^+W^-)$ can $-0.35\%$ around $\mu=100$ GeV. If the cross section of $e^+e^- \to W^+W^-$ can be measured with an accuracy of $0.1\%$, the CEPC will be able to probe the higgsinos with mass up to about 210 GeV. As a comparison, we also present the expected $5\sigma$ sensitivity of LHC search for higgsino world via monojet plus soft dileptons plus $\slashed E_T$ \cite{higgsino-3}. It can be seen that such a direct search strategy may discover the higgsino with mass less than about 205 GeV at 14 TeV LHC with ${\cal L}=1000$ fb$^{-1}$. Therefore, the CEPC may provide a complementary test of the light higgsinos.

\begin{table}[th]
\caption{The effect of currently measured $W$ boson mass ($M_W=80.385 \pm 0.015$ GeV) on the cross section of $e^+e^- \to W^+W^-$ in higgsino world. The benchmark point is: $\mu=152.68$ GeV, $\tan\beta=9.83$, $M_A=2$ TeV, $M_1=1$ TeV, $M_2=2$ TeV, $M_{\rm SUSY}=10$ TeV and $A_t=1.5$ TeV.}
 \vspace*{0.5cm}
\begin{tabular}{|c|c|c|c|}
\hline
$M_W$ (GeV) & $\sigma^{SM}_{\rm one-loop}$ (fb) & $\sigma^{MSSM}_{\rm one-loop}$ (fb) & $\delta$ (\%)\\
\hline
80.385  & 19.37833 & 19.30547 & -0.376\\
\hline
80.415  & 19.4603 & 19.38687 & -0.377\\
\hline
80.355  & 19.29595 & 19.22366 & -0.375\\
\hline
\end{tabular}
\label{mw}
\end{table}

\section{conclusions}\label{section4}
The naturalness criterion applied to supersymmetric theories requires a low higgsino mass, which indicates four nearly degenerate higgsinos with masses $\sim$ 100-300 GeV. The direct way of searching for such a higgsino world from mono-jet events is challenging at high luminosity HL-LHC because of large systematical uncertainties. In this work, I study the indirect effect of higgsinos in the process $e^+e^- \to W^+W^-$ at 240 GeV CEPC. I find that the higgsino mass parameter $\mu \lesssim 210$ GeV can be probed if the accuracy of the measurement of $W^+W^-$ production can reach 0.1\% at the CEPC.

\acknowledgments
This work was supported by the National Natural Science Foundation of China (NNSFC) under grants No. 11705093 and 11305049.


\begin{thebibliography}{99}

\bibitem{mz}
R. Arnowitt and P. Nath, Phys. Rev. D 46, 3981 (1992).




\bibitem{bg}
  R.~Barbieri and G.~F.~Giudice,
  Nucl.\ Phys.\ B {\bf 306}, 63 (1988).

\bibitem{baer-ew-0}
  H.~Baer, V.~Barger, P.~Huang, A.~Mustafayev and X.~Tata,
  Phys.\ Rev.\ Lett.\  {\bf 109}, 161802 (2012)
  [arXiv:1207.3343 [hep-ph]].


\bibitem{nsusy-1}
  C.~Brust, A.~Katz, S.~Lawrence and R.~Sundrum,
  JHEP {\bf 1203}, 103 (2012).

\bibitem{nsusy-2}
  M.~Papucci, J.~T.~Ruderman and A.~Weiler,
  JHEP {\bf1209}, 035 (2012).

\bibitem{nsusy-3}
  L.~J.~Hall, D.~Pinner and J.~T.~Ruderman,
  JHEP {\bf 1204}, 131 (2012);



\bibitem{stop-1}
  H.~Baer, V.~Barger, P.~Huang, D.~Mickelson, A.~Mustafayev, W.~Sreethawong and X.~Tata,
  JHEP {\bf 1312}, 013 (2013)
  Erratum: [JHEP {\bf 1506}, 053 (2015)].


\bibitem{stop-2}
  H.~Baer, V.~Barger, P.~Huang and X.~Tata,
  JHEP {\bf 1205}, 109 (2012).

\bibitem{stop-3}
  J.~Cao, C.~Han, L.~Wu, J.~M.~Yang and Y.~Zhang,
  JHEP {\bf 1211}, 039 (2012).


\bibitem{stop-22}
  C.~Han, F.~Wang and J.~M.~Yang,
  JHEP {\bf 1311}, 197 (2013)
  doi:10.1007/JHEP11(2013)197
  [arXiv:1304.5724 [hep-ph]].

\bibitem{stop-23}
  K.~Kowalska and E.~M.~Sessolo,
  Phys.\ Rev.\ D {\bf 88}, no. 7, 075001 (2013)
  [arXiv:1307.5790 [hep-ph]].

\bibitem{stop-6}
  C.~Han, K.~i.~Hikasa, L.~Wu, J.~M.~Yang and Y.~Zhang,
  JHEP {\bf 1310}, 216 (2013)
  doi:10.1007/JHEP10(2013)216
  [arXiv:1308.5307 [hep-ph]].

\bibitem{stop-7}
  A.~Kobakhidze, N.~Liu, L.~Wu and J.~M.~Yang,
  Phys.\ Rev.\ D {\bf 92}, no. 7, 075008 (2015)
  doi:10.1103/PhysRevD.92.075008
  [arXiv:1504.04390 [hep-ph]].

\bibitem{stop-14}
  M.~Drees and J.~S.~Kim,
  Phys.\ Rev.\ D {\bf 93}, no. 9, 095005 (2016)
  doi:10.1103/PhysRevD.93.095005
  [arXiv:1511.04461 [hep-ph]].

\bibitem{stop-8}
  K.~i.~Hikasa, J.~Li, L.~Wu and J.~M.~Yang,
  Phys.\ Rev.\ D {\bf 93}, no. 3, 035003 (2016)
  doi:10.1103/PhysRevD.93.035003
  [arXiv:1505.06006 [hep-ph]].

\bibitem{stop-9}
  A.~Kobakhidze, N.~Liu, L.~Wu, J.~M.~Yang and M.~Zhang,
  Phys.\ Lett.\ B {\bf 755}, 76 (2016)
  doi:10.1016/j.physletb.2016.02.003
  [arXiv:1511.02371 [hep-ph]].

\bibitem{stop-13}
  J.~S.~Kim, K.~Rolbiecki, R.~Ruiz, J.~Tattersall and T.~Weber,
  Phys.\ Rev.\ D {\bf 94}, no. 9, 095013 (2016)
  doi:10.1103/PhysRevD.94.095013
  [arXiv:1606.06738 [hep-ph]].

\bibitem{stop-10}
  C.~Han, J.~Ren, L.~Wu, J.~M.~Yang and M.~Zhang,
  Eur.\ Phys.\ J.\ C {\bf 77}, no. 2, 93 (2017)
  doi:10.1140/epjc/s10052-017-4662-7
  [arXiv:1609.02361 [hep-ph]].

\bibitem{stop-11}
  G.~H.~Duan, K.~i.~Hikasa, L.~Wu, J.~M.~Yang and M.~Zhang,
  JHEP {\bf 1703}, 091 (2017)
  doi:10.1007/JHEP03(2017)091
  [arXiv:1611.05211 [hep-ph]].

\bibitem{stop-20}
  H.~Baer, V.~Barger, H.~Serce and X.~Tata,
  Phys.\ Rev.\ D {\bf 94}, no. 11, 115017 (2016)
  doi:10.1103/PhysRevD.94.115017
  [arXiv:1610.06205 [hep-ph]].

\bibitem{stop-15}
  F.~Wang, J.~M.~Yang and Y.~Zhang,
  JHEP {\bf 1604}, 177 (2016)
  doi:10.1007/JHEP04(2016)177
  [arXiv:1602.01699 [hep-ph]].

\bibitem{stop-18}
  M.~R.~Buckley, D.~Feld, S.~Macaluso, A.~Monteux and D.~Shih,
  arXiv:1610.08059 [hep-ph].

\bibitem{stop-16}
  H.~Baer, V.~Barger, N.~Nagata and M.~Savoy,
  Phys.\ Rev.\ D {\bf 95}, no. 5, 055012 (2017)
  doi:10.1103/PhysRevD.95.055012
  [arXiv:1611.08511 [hep-ph]].




\bibitem{stop-25}
  G.~G.~Ross, K.~Schmidt-Hoberg and F.~Staub,
  Phys.\ Lett.\ B {\bf 759}, 110 (2016)
  doi:10.1016/j.physletb.2016.05.053
  [arXiv:1603.09347 [hep-ph]].

\bibitem{stop-26}
  G.~G.~Ross, K.~Schmidt-Hoberg and F.~Staub,
  JHEP {\bf 1703}, 021 (2017)
  doi:10.1007/JHEP03(2017)021
  [arXiv:1701.03480 [hep-ph]].

\bibitem{higgs-atlas}
  G.~Aad {\it et al.} [ATLAS Collaboration],
  Phys.\ Lett.\ B {\bf 716}, 1 (2012)
  [arXiv:1207.7214 [hep-ex]].



\bibitem{higgs-cms}
  S.~Chatrchyan {\it et al.} [CMS Collaboration],
  Phys.\ Lett.\ B {\bf 716}, 30 (2012)
  doi:10.1016/j.physletb.2012.08.021
  [arXiv:1207.7235 [hep-ex]].

\bibitem{run2-stop}
The CMS collaboration, CMS-PAS-SUS-16-029.

\bibitem{run2-gluino}
The ATLAS collaboration, ATLAS-CONF-2015-067.



\bibitem{nath}
  K.~L.~Chan, U.~Chattopadhyay and P.~Nath,
  Phys.\ Rev.\ D {\bf 58}, 096004 (1998)
  doi:10.1103/PhysRevD.58.096004
  [hep-ph/9710473].

\bibitem{feng}
  J.~L.~Feng, K.~T.~Matchev and T.~Moroi,
  Phys.\ Rev.\ Lett.\  {\bf 84}, 2322 (2000)
  doi:10.1103/PhysRevLett.84.2322
  [hep-ph/9908309].

\bibitem{kane}
  G.~L.~Kane,
  Nucl.\ Phys.\ Proc.\ Suppl.\  {\bf 62}, 144 (1998)
  doi:10.1016/S0920-5632(97)00651-8
  [hep-ph/9709318].

\bibitem{higgsino-1}
  C.~Han, A.~Kobakhidze, N.~Liu, A.~Saavedra, L.~Wu and J.~M.~Yang,
  JHEP {\bf 1402}, 049 (2014)
  doi:10.1007/JHEP02(2014)049
  [arXiv:1310.4274 [hep-ph]].

\bibitem{higgsino-2}
  Z.~Han, G.~D.~Kribs, A.~Martin and A.~Menon,
  Phys.\ Rev.\ D {\bf 89}, 075007 (2014);

\bibitem{higgsino-3}
  H.~Baer, A.~Mustafayev and X.~Tata,
  Phys.\ Rev.\ D {\bf 90}, 115007 (2014);

\bibitem{higgsino-4}
  P.~Schwaller and J.~Zurita,
  JHEP {\bf 1403}, 060 (2014);

\bibitem{higgsino-5}
  D.~Barducci,  {\it et al.},
  arXiv:1504.02472 [hep-ph];


\bibitem{ww-sm-1}
  W.~Beenakker, K.~Kolodziej and T.~Sack,
  Phys.\ Lett.\ B {\bf 258}, 469 (1991).
  doi:10.1016/0370-2693(91)91120-K


\bibitem{ww-sm-2}
  Z.~Hioki,
  Nucl.\ Phys.\ B {\bf 316}, 1 (1989).
  doi:10.1016/0550-3213(89)90383-0

\bibitem{ww-sm-3}
  S.~Dittmaier, M.~Bohm and A.~Denner,
  Nucl.\ Phys.\ B {\bf 376}, 29 (1992)
  Erratum: [Nucl.\ Phys.\ B {\bf 391}, 483 (1993)].
  doi:10.1016/0550-3213(92)90066-K, 10.1016/0550-3213(93)90156-J

\bibitem{ww-sm-4}
  J.~Fleischer, K.~Kolodziej and F.~Jegerlehner,
  Phys.\ Rev.\ D {\bf 47}, 830 (1993).
  doi:10.1103/PhysRevD.47.830

\bibitem{ww-sm-5}
  J.~Fleischer, F.~Jegerlehner, K.~Kolodziej and G.~J.~van Oldenborgh,
  Comput.\ Phys.\ Commun.\  {\bf 85}, 29 (1995)
  doi:10.1016/0010-4655(94)00113-G
  [hep-ph/9405380].

\bibitem{ww-sm-6}
  A.~Denner, S.~Dittmaier, M.~Roth and D.~Wackeroth,
  Nucl.\ Phys.\ B {\bf 587}, 67 (2000)
  doi:10.1016/S0550-3213(00)00511-3
  [hep-ph/0006307].

\bibitem{ww-mssm-1}
  S.~Alam,
  Phys.\ Rev.\ D {\bf 50}, 124 (1994);
  Phys.\ Rev.\ D {\bf 50}, 148 (1994);
  Phys.\ Rev.\ D {\bf 50}, 174 (1994).

\bibitem{ww-mssm-2}
  S.~Alam, K.~Hagiwara, S.~Kanemura, R.~Szalapski and Y.~Umeda,
  Phys.\ Rev.\ D {\bf 62}, 095011 (2000)
  doi:10.1103/PhysRevD.62.095011
  [hep-ph/0002066].

\bibitem{ww-mssm-3}
  K.~Hagiwara, S.~Kanemura, M.~Klasen and Y.~Umeda,
  Phys.\ Rev.\ D {\bf 68}, 113011 (2003)
  doi:10.1103/PhysRevD.68.113011
  [hep-ph/0212135].

\bibitem{ww-mssm-4}
  A.~A.~Barrientos Bendezu, K.~P.~O.~Diener and B.~A.~Kniehl,
  Phys.\ Lett.\ B {\bf 478}, 255 (2000)
  doi:10.1016/S0370-2693(00)00259-8
  [hep-ph/0002058].

\bibitem{ww-mssm-5}
  T.~Hahn,
  Nucl.\ Phys.\ B {\bf 609}, 344 (2001)
  doi:10.1016/S0550-3213(01)00300-5
  [hep-ph/0007062].

\bibitem{cepc-0}
CEPC-SppC Preliminary Conceptual Design Report, http://cepc.ihep.ac.cn/preCDR/volume.html

\bibitem{cepc-1}
  M.~McCullough,
  Phys.\ Rev.\ D {\bf 90}, no. 1, 015001 (2014)
  Erratum: [Phys.\ Rev.\ D {\bf 92}, no. 3, 039903 (2015)]
  doi:10.1103/PhysRevD.90.015001, 10.1103/PhysRevD.92.039903
  [arXiv:1312.3322 [hep-ph]].

\bibitem{cepc-2}
  C.~Shen and S.~h.~Zhu,
  Phys.\ Rev.\ D {\bf 92}, no. 9, 094001 (2015)
  doi:10.1103/PhysRevD.92.094001
  [arXiv:1504.05626 [hep-ph]].

\bibitem{cepc-3}
  F.~P.~Huang, P.~H.~Gu, P.~F.~Yin, Z.~H.~Yu and X.~Zhang,
  Phys.\ Rev.\ D {\bf 93}, no. 10, 103515 (2016)
  doi:10.1103/PhysRevD.93.103515
  [arXiv:1511.03969 [hep-ph]].

\bibitem{cepc-4}
  A.~Kobakhidze, N.~Liu, L.~Wu and J.~Yue,
  Phys.\ Rev.\ D {\bf 95}, no. 1, 015016 (2017)
  doi:10.1103/PhysRevD.95.015016
  [arXiv:1610.06676 [hep-ph]].

\bibitem{cepc-5}
  J.~Fan, M.~Reece and L.~T.~Wang,
  JHEP {\bf 1508}, 152 (2015)
  doi:10.1007/JHEP08(2015)152
  [arXiv:1412.3107 [hep-ph]].

\bibitem{cepc-6}
  Q.~H.~Cao, H.~R.~Wang and Y.~Zhang,
  Chin.\ Phys.\ C {\bf 39}, no. 11, 113102 (2015)
  doi:10.1088/1674-1137/39/11/113102
  [arXiv:1505.00654 [hep-ph]].

\bibitem{cepc-7}
  S.~L.~Hu, N.~Liu, J.~Ren and L.~Wu,
  J.\ Phys.\ G {\bf 41}, no. 12, 125004 (2014)
  doi:10.1088/0954-3899/41/12/125004
  [arXiv:1402.3050 [hep-ph]].

\bibitem{cepc-8}
  S.~Gori, J.~Gu and L.~T.~Wang,
  JHEP {\bf 1604}, 062 (2016)
  doi:10.1007/JHEP04(2016)062
  [arXiv:1508.07010 [hep-ph]].

\bibitem{cepc-9}
  Q.~H.~Cao, Y.~Li, B.~Yan, Y.~Zhang and Z.~Zhang,
  Nucl.\ Phys.\ B {\bf 909}, 197 (2016)
  doi:10.1016/j.nuclphysb.2016.05.010
  [arXiv:1604.07536 [hep-ph]].

\bibitem{cepc-10}
  S.~F.~Ge, H.~J.~He and R.~Q.~Xiao,
  JHEP {\bf 1610}, 007 (2016)
  doi:10.1007/JHEP10(2016)007
  [arXiv:1603.03385 [hep-ph]].

\bibitem{cepc-11}
  C.~Cai, Z.~H.~Yu and H.~H.~Zhang,
  arXiv:1611.02186 [hep-ph].

\bibitem{cepc-12}
  L.~Bian, J.~Shu and Y.~Zhang,
  JHEP {\bf 1509}, 206 (2015)
  doi:10.1007/JHEP09(2015)206
  [arXiv:1507.02238 [hep-ph]].

\bibitem{cepc-13}
  J.~Cao, C.~Han, J.~Ren, L.~Wu, J.~M.~Yang and Y.~Zhang,
  Chin.\ Phys.\ C {\bf 40}, no. 11, 113104 (2016)
  doi:10.1088/1674-1137/40/11/113104
  [arXiv:1410.1018 [hep-ph]].

\bibitem{cepc-14}
  C.~Han, A.~Kobakhidze, N.~Liu, L.~Wu and B.~Yang,
  Nucl.\ Phys.\ B {\bf 890}, 388 (2014)
  doi:10.1016/j.nuclphysb.2014.11.021
  [arXiv:1405.1498 [hep-ph]].

\bibitem{cepc-15}
  N.~Liu, J.~Ren, L.~Wu, P.~Wu and J.~M.~Yang,
  JHEP {\bf 1404}, 189 (2014)
  doi:10.1007/JHEP04(2014)189
  [arXiv:1311.6971 [hep-ph]].

\bibitem{on-shell}
  M.~Bohm, H.~Spiesberger and W.~Hollik,
  Fortsch.\ Phys.\ 34 (1986) 687.

\bibitem{soft-photon}
S. Dawson and L. Reina, Phys. Rev. D59, 054012 (1999).

\bibitem{vegas}
G.P. Legage, J. Comput. Phys. 27, 192(1978).


\bibitem{feynarts}
T. Hahn, Comput. Phys. Commun. \textbf{140}, 418 (2001).


\bibitem{formcalc}
T. Hahn, M. Perez-Victoria, Comput. Phys. Commun. \textbf{118}, 153
(1999).

\bibitem{looptools}
G. J. van Oldenborgh, Phys Commun \textbf{66}, 1 (1991).

\bibitem{cepc-ew}
https://indico.cern.ch/event/432527/contributions/1071856/attachments/1321305/1981584/ICHEP2016zhijunv2.pdf

\end{thebibliography}
\end{document}